\documentclass[11pt,twoside]{article}
\usepackage{asp2010}

\resetcounters

\begin{document}

\title{Fundamental Properties of Young Binary Stars}
\author{Gail H. Schaefer$^1$}
\affil{$^1$The CHARA Array of Georgia State University, Mount Wilson Observatory, Mount Wilson, CA 91023}

\begin{abstract}
Spatially resolving the orbits of double-lined spectroscopic binaries provides a way to measure the dynamical masses of the component stars.  At the distances to the nearest star forming regions, high angular resolution techniques are required to resolve these short period systems.  In this paper, I provide an overview of the few low-mass pre-main-sequence spectroscopic binaries that have been resolved thus far using long-baseline optical/infrared interferometers.  I also compiled a list of known spectroscopic binaries in nearby star forming regions (Taurus, Orion, Ophiuchus, Scorpius-Centaurus, etc.) and show that with modest improvements in the sensitivity of interferometers with 200-300 meter baselines, we can build a significant set of pre-main sequence stars with precise mass determinations.  This is important for validating and distinguishing among the theoretical calculations of evolution for young stars.  
\end{abstract}

\section{Introduction}

\begin{figure}[!t]
\begin{center}
  \plotone{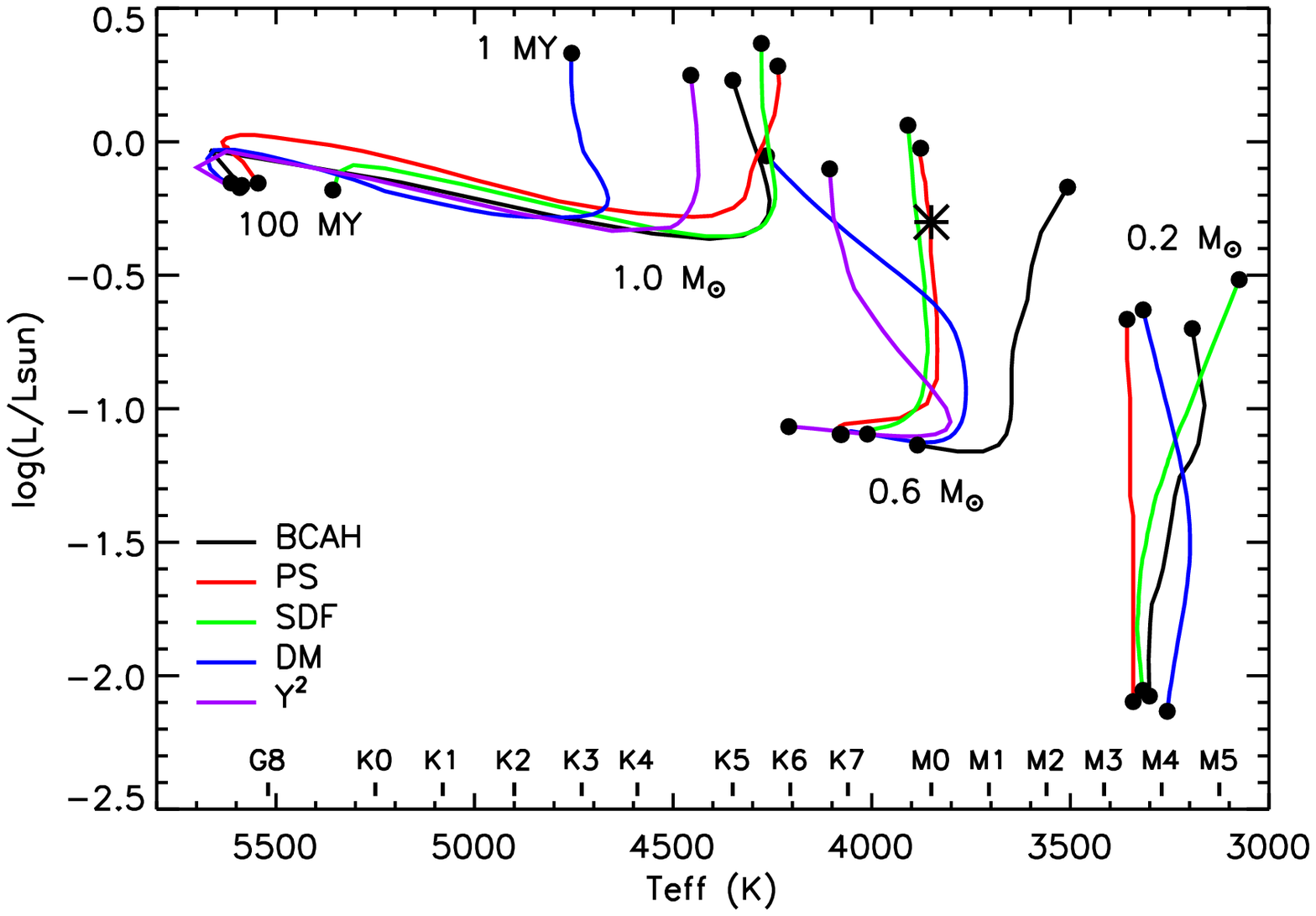}
  \caption{Pre-main-sequence evolutionary tracks computed by \citet[][DM]{dantona97}, \citet[][BCAH]{baraffe98}, \citet[][PS]{palla99}, \citet[][SDF]{siess00}, and \citet[][Y$^2$]{yi03}.  The tracks are for masses of 0.2, 0.6, and 1.0 $M_\odot$ and correspond to ages from 1 Myr to 100 Myr.  
We indicate spectral types based on the effective temperature relations used in \citet{kenyon95} for G8-M0 and the T-Tauri temperature scale defined in \citet{luhman03} for M1-M7.  The asterisk shows the location of an M0 star with 0.5 L$_\odot$.\label{fig.hrdiag}}   
\end{center}
\end{figure}

Determining precise masses and ages of pre-main-sequence (PMS) stars is important for understanding how stars form and evolve over time.  Different theoretical models of stellar evolution predict a large range of stellar masses for a given effective temperature, luminosity, and metallicity at young ages.  This is particularly the case for low-mass PMS stars \citep[M $<$ 1 M$_\odot$; e.g.][]{simon01,hillenbrand04,mathieu07}.  Discrepancies between the models arise from differences in the treatment of interior convection, atmospheric opacities, and initial conditions \citep[e.g.][]{allard97, chabrier00, baraffe03}.  Figure~\ref{fig.hrdiag} illustrates the discrepancies among different evolutionary models.  Measuring precise masses of young stars is required for testing the different sets of evolutionary tracks.

\begin{figure}[!t]
\begin{center}
  \plotone{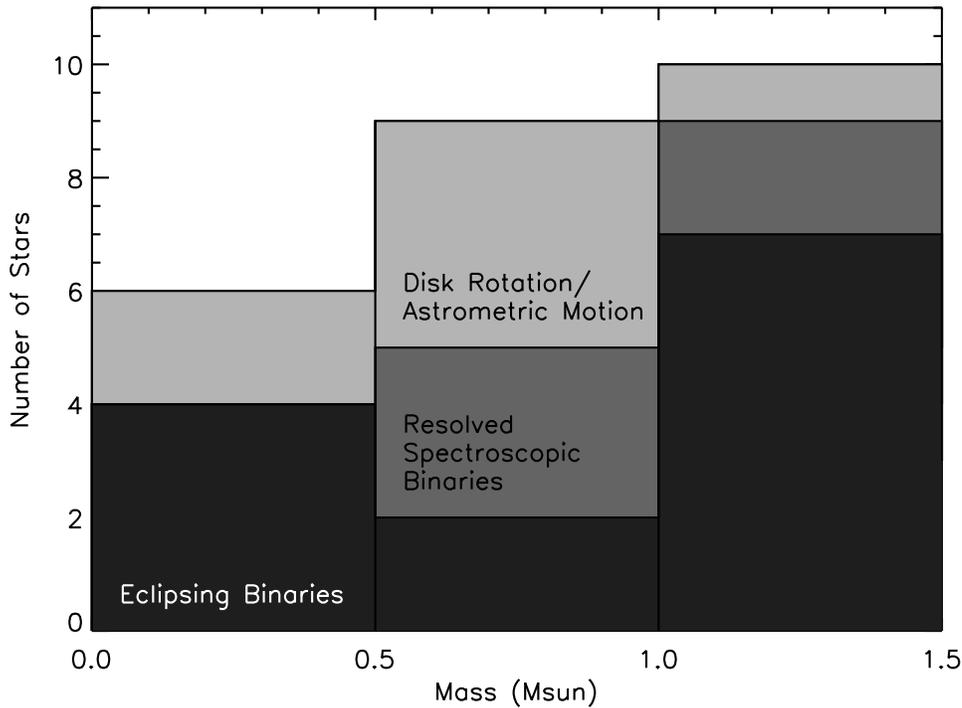}
\caption{Histogram of dynamical mass measurements of low-mass stars in nearby star forming regions (Taurus, Orion, Ophiuchus, Scorpius-Centaurus, TW Hydra) with internal precision measured to better than $\sim$ 10\%.  The shaded bins indicate the method of mass measurement: eclipsing spectroscopic binaries (black), spatially resolved spectroscopic binaries (dark grey), and Keplerian rotation of circumstellar disks and astrometric center of mass motion in visual binaries (light grey).  Note that the masses in the last category are dependent on the distance to the star.  References for eclipsing SB: \citet{popper87}, \citet{alencar03}, \citet{covino04}, \citet{stassun04}, \citet{irwin07}, \citet{cargile08}, \citet{stempels08}; resolved SB: \citet{steffen01}, \citet{boden05}; disk rotation: \citet{simon00}, \citet{prato02}; and astromery: \citet{schaefer10}.\label{fig.histo}}.
\end{center}
\end{figure}

Stellar masses can be measured through the following dynamical techniques: (1) observations of eclipsing double-lined spectroscopic binaries where the masses and radii of the component stars are determined from the radial velocity variations and timing of the eclipse light curve (Torres summarized results from eclipsing binaries at this conference), (2) spatially resolving the visual orbit of a double-lined spectroscopic binary to determine the component stellar masses and distance to the system, (3) measuring the astrometric motion of component stars around their center of mass \citep[e.g. the triple system Elias 12;][]{schaefer10}, (4) mapping the Keplerian rotation of circumstellar disks to determine the mass of the central star \citep{guilloteau98,simon00}.  The last two techniques require an accurate parallax to scale the masses to the proper distance.  Figure~\ref{fig.histo} shows a histogram of the current sample of dynamical masses of PMS stars measured using these techniques.

In this paper I focus on spatially resolving double-lined spectroscopic binaries to determine the masses of PMS stars.  In addition to masses, a comparison of the physical size of the projected semi-major axis from the spectroscopic orbit ($a\sin{i}$ in AU) with the angular scale ($a$ in mas) and inclination $i$ from the visual orbit provide an orbital parallax for the system.  Accurate distances are necessary for determining the absolute magnitudes and luminosities for plotting the stars on the HR diagram to compare with evolutionary tracks.  Additionally, distances for independent systems will aid in mapping the three-dimensional structure of nearby star-forming regions \citep{loinard08,torres09}.  The distribution of masses, mass ratios, and orbital parameters of PMS binaries across a variety of star forming regions will help test theories of binary star formation \citep{ballesteros07,bonnell07}.

\section{Interferometric Observations of Young Binary Stars}

Double-lined spectroscopic binaries tend to sample short-period systems ($P < $ 10 years, corresponding to velocity amplitudes $<$ 5$-$10 km/s).  For a circular binary with equal mass components and a total mass of 1 M$_\odot$, a period of 10 years corresponds to a semi-major axis of 4.6 AU or $\sim$ 33 mas for an edge-on system scaled to the distance of the Taurus star forming region ($d \sim$ 140 pc).  The diffraction limit ($1.22\lambda/D$) of a 10-meter telescope in the $H$-band is $\sim$~40~mas.  Therefore, binaries that can be spatially resolved through adaptive optics imaging, speckle interferometry, or aperture masking using large aperture telescopes in the infrared or with the Hubble Space Telescope in the optical have periods measured on the order of decades or longer \citep[e.g.][]{schaefer06}.  Clearly, long-baseline optical/infrared interferometry is required to spatially resolve the short period spectroscopic binaries in nearby star-forming regions.

\begin{figure}[!t]
\begin{center}
  \plotone{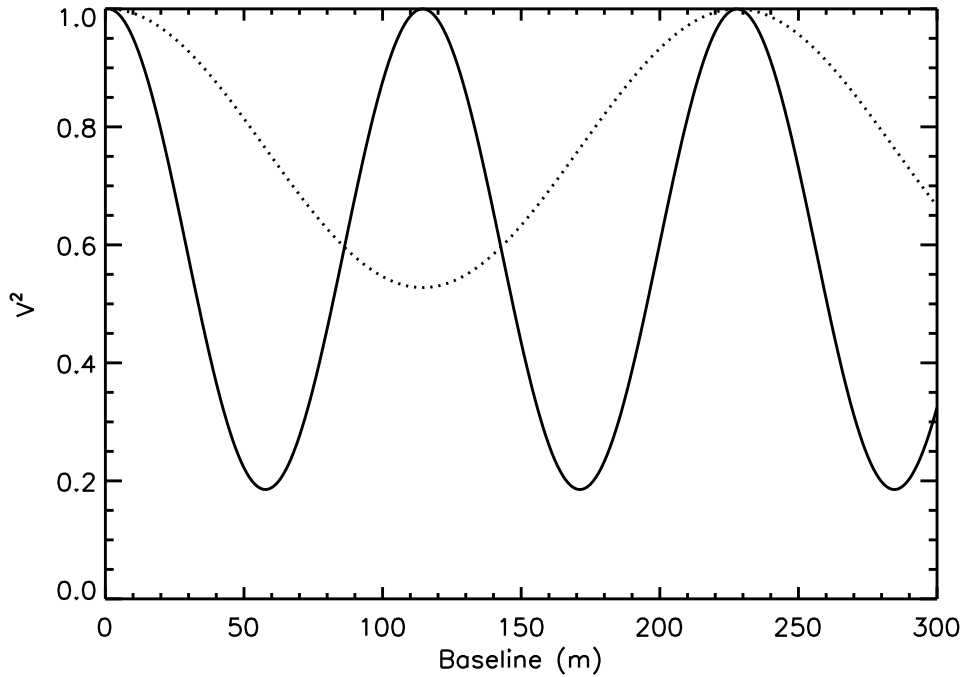}
\caption{Example visibility curves of a binary star in the $K$-band for interferometer baselines out to 300 meters.  The solid line shows a 4.0 mas binary system with a flux of $\Delta K$ = 1.0, while the dotted line shows a 2.0 mas binary with a flux ratio of $\Delta K$ = 2.0.  In both cases the angular diameters of the component stars are unresolved.\label{fig.vis}}
\end{center}
\end{figure}

The PMS double-lined spectroscopic binary NTT 045251$+$3016 was spatially resolved using the Fine Guidance Sensors on the Hubble Space Telescope by \citet{steffen01}.  At smaller separations, long-baseline interferometry becomes a critical tool for resolving young binary systems.  Interferometers measure the fringe contrast or visibility of the source.  For a binary star, the visibility is a periodic function given by:
\begin{equation}
V^2_{\rm bin} = \frac{V^2_1 + f^2V^2_2 + 2f|V_1||V_2|\cos{(2\pi(u\Delta \alpha + v\Delta \beta))}}{(1 + f)^2}
\end{equation}
where $(\Delta \alpha, \Delta \beta)$ represent the separation of the binary in right ascension and declination, ($u,v$) are the spatial frequencies of the interferometer baseline projected on the sky, $f$ is the binary flux ratio, and $V_1$ and $V_2$ are the uniform disk visibilities of the component stars \citep[e.g.][]{boden99}.  The spacing between the peaks in the visibility curve gives the separation of the binary projected along the interferometer baseline while the minimum indicates the flux ratio (see Figure~\ref{fig.vis}).  Additionally, for interferometers that combine the light from three or more telescopes, closure phases can also be used to determine the binary separation and flux ratio \citep{monnier07}.  The Keck Interferometer was used to spatially resolve the orbits of three PMS double-lined spectroscopic binaries: HD 98800 B \citep{boden05}, V773 Tau A \citep{boden07}, and Haro 1-14c \citep{schaefer08}.

\section{Prospects for the Future}

\begin{figure}[!t]
\begin{center}
   \plotone{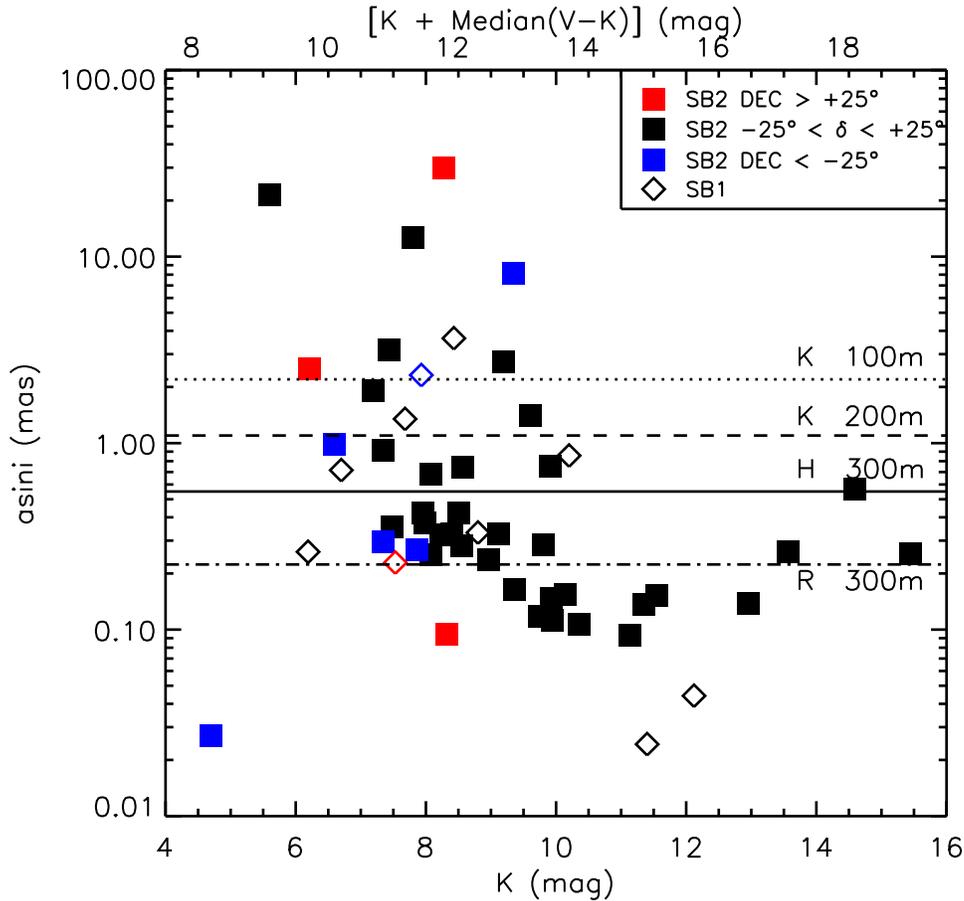}
\caption{Distribution of projected semi-major axes of low-mass PMS spectroscopic binaries in nearby star forming regions.  The semi-major axis has been scaled by the average distance to the star forming region where the star is located.  The squares indicate double-lined spectroscopic binaries.  Red indicates targets at northern declinations, blue southern declinations, and black declinations accessible from interferometers in both hemispheres.  The open diamonds represent $a_1\sin{i}$ for single-lined spectroscopic binaries.  The bottom axis plots the $K$-band magnitude while the top axis shows a proxy for the $V$-band magnitude (the individual $K$-band magnitude of the binary plus the median ($V-K$) for all stars plotted).  The horizontal lines represent the angular resolution ($\frac{\lambda}{2B}$) for different interferometer baseline lengths and filter wavelengths (dotted: $K$-band, 100 m; dashed: $K$-band, 200 m; solid: $H$-band, 300 m; dash-dotted: $R$-band, 300 m).\label{fig.aK}}
\end{center}
\end{figure}

Table~\ref{tab.sb} provides a summary of known double-lined and single-lined spectroscopic binaries in nearby star forming regions (Taurus-Auriga, Orion, Ophiuchus, Scorpius-Centaurus, Lupus, TW Hydra, NGC 2264).  Additional binary candidates, including some with longer periods ($P >$ 100 days), are identified in \citet{melo03}, \citet{guenther07}, \citet{prato07}, and \citet{tobin09}.  In Figure~\ref{fig.aK}, I use the sample of binaries in Table~\ref{tab.sb} to plot the projected semi-major axis in angular units versus the $K$-band magnitude.  This plot shows a lack of known PMS spectroscopic binaries at long periods (large separations) and faint magnitudes ($K >$ 10 mag).  As the sensitivity of long-baseline interferometers improve, it will be useful to fill in this region of parameter space.  In addition to spectroscopic surveys, Gordon et al. discussed at this conference the feasibility of an interferometric survey to search for PMS binary stars.

The horizontal lines in Figure~\ref{fig.aK} show the angular resolution of 100-300 meter interferometric baselines operating at the $K$, $H$, or $R$-bands.  Typical limiting magnitudes for the CHARA array (30-330 meter baselines) are $V \sim 10$ mag for tip/tilt guiding and $K \sim 7$ mag for fringe tracking.  This is right at the edge of the distribution of PMS binaries; a few of the fainter targets might be accessible in favorable seeing conditions.  With upgrades to improve the tip/tilt system and plans to add full adaptive optics to the CHARA telescopes, expected limits of $V \sim 16$ mag and $K \sim 10$ mag could bring a large portion of the known population within reach of CHARA.  Likewise, the current sensitivity for the Very Large Telescope Interferometer (8-200 meter baselines) is $V \sim 13.5$ mag and $H \sim 7$ mag.  The Magdalena Ridge Observatory Interferometer that is under construction (8-340 meter baselines) is expected to achieve a limiting magnitude of $H \sim 14$ mag.  With combined efforts to push spectroscopic surveys for longer period binaries fainter and with the expected improvements in sensitivity of long-baseline optical/IR interferometers, we can continue assembling a large sample of low-mass PMS stars with precision masses to calibrate the evolutionary tracks.

\acknowledgements I would like to thank M. Simon for reading through a draft of this paper.

\bibliography{schaefer_youngbinary}

\begin{table}[!ht]
\caption{Sample of Spectroscopic Binaries for T Tauri Stars ($M <$ 1.5 M$_\odot$)}
\smallskip
\begin{center}
{\scriptsize
\begin{tabular}{lllcccrl}
\tableline
\noalign{\smallskip}
\multicolumn{8}{c}{Double-lined Spectroscopic Binaries} \\
Name & RA   & DEC       & SpT & V     & K     & P\hspace{1em}    & Reference \\
     & (hr) & \hspace{0.5em}($^\circ$) &     & (mag) & (mag) & (days) &                    \\ 
\noalign{\smallskip}
\tableline
\noalign{\smallskip}
          AK Sco$^*$ &  16 & $-$36 &       F5 &  8.8 &  6.6 &     13.6 &    \citet{alencar03}   \\
ASAS J052821+0338.5$^*$ & 05 & $+$03 &  K1/K3 & 11.2 &  9.0 &      3.9 &    \citet{stempels08}  \\
              Cru 3  &  12 & $-$59 &    K5/K5 & 10.8 &  7.5 &     58.3 &    \citet{alcala02}    \\   
              DQ Tau &  04 & $+$17 &       M0 & 13.7 &  8.0 &     15.8 &    \citet{mathieu97}   \\
        EK Cep$^{*}$ &  21 & $+$69 &     A1/G &  7.9 &  7.6 &      4.4 &    \citet{tomkin83}    \\
     GSC 06213-00306 &  06 & $-$00 &       K1 & 10.9 &  7.4 &    166.9 &    \citet{guenther07}  \\
   Haro 1-14c$^\dag$ &  16 & $-$24 &  K5/M1.5 & 12.3 &  7.8 &    592.1 &    \citet{schaefer08}  \\
            HD 34700 &  05 & $+$05 &    G0IVe &  9.2 &  7.5 &     23.5 &    \citet{torres04}    \\
   HD 98800 B$^\dag$ &  11 & $-$24 &       K5 &  9.1 &  5.6 &    314.3 &    \citet{boden05}     \\
           HD 155555 &  17 & $-$66 &       G5 &  6.7 &  4.7 &      1.7 &    \citet{pasquini91}  \\
        JW 380$^{*}$ &  05 & $-$05 &       M5 & 17.2 & 11.1 &      5.3 &    \citet{irwin07}  \\
          MML 53$^*$ &  14 & $-$35 &    K2IVe & 11.3 &  7.9 &      2.1 &    \citet{hebb10} \\
NTTS 045251+3016$^\dag$ & 04 & $+$30 &     K7 & 11.6 &  8.3 &   2525.0 &    \citet{steffen01}   \\
    NTTS 155913-2233 &  15 & $-$22 &       K5 & 11.2 &  8.1 &      2.4 &    \citet{prato02}   \\
    NTTS 160905-1859 &  16 & $-$18 &       K1 & 11.7 &  8.1 &     10.4 &    \citet{prato02}   \\
    NTTS 162814-2427 &  16 & $-$24 &       K7 & 12.2 &  7.2 &     36.0 &    \citet{mathieu89}   \\
       Parenago 1540 &  05 & $-$05 &       K3 & 11.3 &  7.9 &     33.7 &    \citet{marschall88} \\
       Parenago 1771 &  05 & $-$05 &       K4 & 13.5 &  9.6 &    149.5 &    \citet{prato02}   \\
   Parenago 1802$^*$ &  05 & $-$05 &       M2 & 15.4 &  9.9 &      4.7 &    \citet{cargile08}   \\
       Parenago 1925 &  05 & $-$05 &       K3 & 13.4 &  8.5 &     32.9 &    \citet{prato02}   \\
       Parenago 2486 &  05 & $-$05 &       G5 & 11.4 &  9.9 &      5.2 &    \citet{mathieu94}  \\
       Parenago 2494 &  05 & $-$06 &       K0 & 10.8 &  8.4 &     19.5 &    \citet{reipurth02}  \\
            ROXR1 14 &  16 & $-$24 &       M1 & 16.5 &  9.1 &      5.7 &    \citet{rosero11}    \\
     RX J0350.5-1355 &  03 & $-$13 &    K0/K1 & 10.7 & 15.4 &      9.3 &    \citet{covino01}    \\
     RX J0441.0-0839 &  04 & $-$08 &    G3/G8 & 10.4 &  8.6 &     13.6 &    \citet{covino01}    \\
     RX J0528.9+1046 &  05 & $+$10 &       K0 & 10.7 & 10.1 &      7.7 &    \citet{torres02}    \\
     RX J0529.3+1210 &  05 & $+$12 &    K7-M0 & 12.9 &  9.2 &    461.9 &    \citet{mace09}      \\
 RX J0529.4+0041$^*$ &  05 & $+$00 &    K1/K7 & 12.3 &  9.7 &      3.0 &    \citet{covino04}    \\
     RX J0530.7-0434 &  05 & $-$04 &    K2/K2 & 11.5 &  8.6 &     40.5 &    \citet{covino01}    \\
     RX J0532.1-0732 &  05 & $-$07 &    K2/K3 & 12.6 &  9.9 &     46.9 &    \citet{covino01}    \\
     RX J0541.4-0324 &  05 & $-$03 &    G8/K3 & 11.8 &  9.4 &      5.0 &    \citet{covino01}    \\
     RX J1559.2-3814 &  15 & $-$38 &     M1.5 & 13.6 &  9.3 &    474.0 &    \citet{guenther07}  \\
     RX J1622.7-2325 &  16 & $-$23 &       M1 & 15.7 &  8.2 &      3.2 &    \citet{rosero11}    \\
                 S29 &  05 & $-$02 &     K9.5 & 14.2 & 11.6 &      8.7 &    \citet{sacco08}     \\
                 S53 &  05 & $-$02 &     M4.5 & 17.4 & 11.3 &      8.5 &    \citet{sacco08}     \\
                 S84 &  05 & $-$02 &     M2.5 & 17.8 & 12.9 &      6.1 &    \citet{sacco08}     \\
                 S85 &  05 & $-$02 &     M1.0 & 18.3 & 13.6 &     12.8 &    \citet{sacco08}     \\
                 S89 &  05 & $-$02 &     K2.0 & 16.5 & 14.6 &     13.8 &    \citet{sacco08}     \\
            UZ Tau E &  04 & $+$25 &       M2 & 11.2 &  7.3 &     19.0 &    \citet{martin05}    \\
       V1174 Ori$^*$ &  05 & $-$05 &       M0 & 14.3 & 10.3 &      2.6 &    \citet{stassun04}   \\
           V4046 Sgr &  18 & $-$32 &       K5 & 10.4 &  7.3 &      2.4 &    \citet{stempels04}  \\
            V773 Tau &  04 & $+$28 &       K2 & 10.7 &  6.2 &     51.1 &    \citet{boden07}     \\
            V826 Tau &  04 & $+$32 &       K7 & 12.1 &  8.3 &      3.9 &    \citet{reipurth90}  \\
          Walker 134 &  06 & $+$09 &       G5 & 12.4 &  9.8 &      6.4 &    \citet{padgett94}   \\
\noalign{\smallskip}
\tableline
\noalign{\smallskip}
\multicolumn{8}{c}{Single-lined Spectroscopic Binaries} \\
Name & RA   & DEC       & SpT & V     & K     & P\hspace{1em}   & Reference \\
     & (hr) & \hspace{0.5em}($^\circ$) &     & (mag) & (mag) & (days) &                   \\ 
\noalign{\smallskip}
\tableline
\noalign{\smallskip}
              GW Ori &  05 & $+$11 &     G5   &  9.8 &  6.2 &    241.9 &    \citet{mathieu91}  \\
              LkCa 3 &  04 & $+$27 &     M1   & 12.1 &  7.5 &     12.9 &    \citet{mathieu94}  \\  
     GSC 06209-00735 &  06 & $-$00 &     K2   & 11.0 &  8.4 &   2045.0 &    \citet{guenther07} \\
    NTTS 155808-2219 &  15 & $-$22 &     M3   & 13.7 &  8.8 &     16.9 &    \citet{mathieu94}  \\
    NTTS 160814-1857 &  16 & $-$18 &     K2   & 11.9 &  7.7 &    144.7 &    \citet{mathieu89}  \\
   NTTS 162819-2423S &  16 & $-$24 &     G8   & 10.6 &  6.7 &     89.1 &    \citet{mathieu89}  \\
     RX J1220.6-7539 &  12 & $-$75 &     K2   & 10.7 &  7.9 &    613.9 &    \citet{guenther07} \\
                 S96 &  05 & $-$02 &     --   & 15.8 & 12.1 &      3.9 &    \citet{sacco08}    \\
             VSB 111 &  06 & $+$09 &     G8   & 12.4 & 10.2 &    879.0 &    \citet{mathieu94}  \\  
             VSB 126 &  06 & $+$09 &     K0   & 13.4 & 11.4 &     12.9 &    \citet{mathieu94}  \\  
\noalign{\smallskip}
\tableline \\
\end{tabular}
}
\end{center}
\vspace{-0.7em}
{\scriptsize $^*$ Eclipsing spectroscopic binary. \\
$^\dag$ Spatially resolved spectroscopic binary. }
\label{tab.sb}
\end{table}

\end{document}